\pdfoutput=1

\documentclass[9pt,twocolumn,twoside]{newclass}

\title{DeepSTORM3D: dense three dimensional localization microscopy and point spread function design by deep learning}

\date{}

\author[1,2]{Elias Nehme}
\author[3]{Daniel Freedman}
\author[2]{Racheli Gordon}
\author[2,4]{Boris Ferdman}
\author[2]{Lucien E. Weiss}
\author[2]{Onit Alalouf}
\author[2,4]{Reut Orange}
\author[1]{Tomer Michaeli}
\author[2,4,*]{Yoav Shechtman}
\affil[1]{Department of Electrical Engineering, Technion, 32000 Haifa, Israel}
\affil[2]{Department of Biomedical Engineering \(\&\) Lorry I. Lokey Center for \protect\\ Life Sciences and Engineering, Technion, 32000 Haifa, Israel}
\affil[3]{Google Research, Haifa, Israel}
\affil[4]{Russel Berrie Nanotechnology Intitute, Technion, 32000 Haifa, Israel}
\affil[*]{Corresponding author: yoavsh@bm.technion.ac.il}

\begin{document}
  
\maketitle
    
\begin{abstract}
Localization microscopy is an imaging technique in which the positions of individual nanoscale point emitters (e.g. fluorescent molecules) are determined at high precision from their images. This is the key ingredient in single/multiple-particle-tracking and several super-resolution microscopy approaches. Localization in three-dimensions (3D) can be performed by modifying the image that a point-source creates on the camera, namely, the point-spread function (PSF). The PSF is engineered using additional optical elements to vary distinctively with the depth of the point-source. However, localizing multiple adjacent emitters in 3D poses a significant algorithmic challenge, due to the lateral overlap of their PSFs. Here, we train a neural network to receive an image containing densely overlapping PSFs of multiple emitters over a large axial range, and output a list of their 3D positions. Furthermore, we then use the network to design the optimal PSF for the multi-emitter case. We demonstrate our approach numerically as well as experimentally by 3D STORM imaging of mitochondria, and volumetric imaging of dozens of fluorescently-labeled telomeres occupying a mammalian nucleus in a single snapshot.
\end{abstract}

\section{Introduction}

Determining the nanoscale positions of point emitters forms the basis of localization microscopy techniques such as single particle tracking \citep{katayama2009real, manzo2015review}, (fluorescence) photoactivated localization microscopy (f)PALM \citep{Betzig,HESS20064258}, stochastic optical reconstruction microscopy (STORM) \citep{Rust2006}, and related single molecule localization microscopy (SMLM) methods. These techniques have revolutionized biological imaging, revealing cellular processes and structures at the nanoscale \citep{SAHL2013778}. Notably, most samples of interest extend in three dimensions, necessitating three-dimensional (3D) localization microscopy \citep{von2017three}.

\begin{figure}[!ht]
\centering
\includegraphics[width=0.335\textwidth]{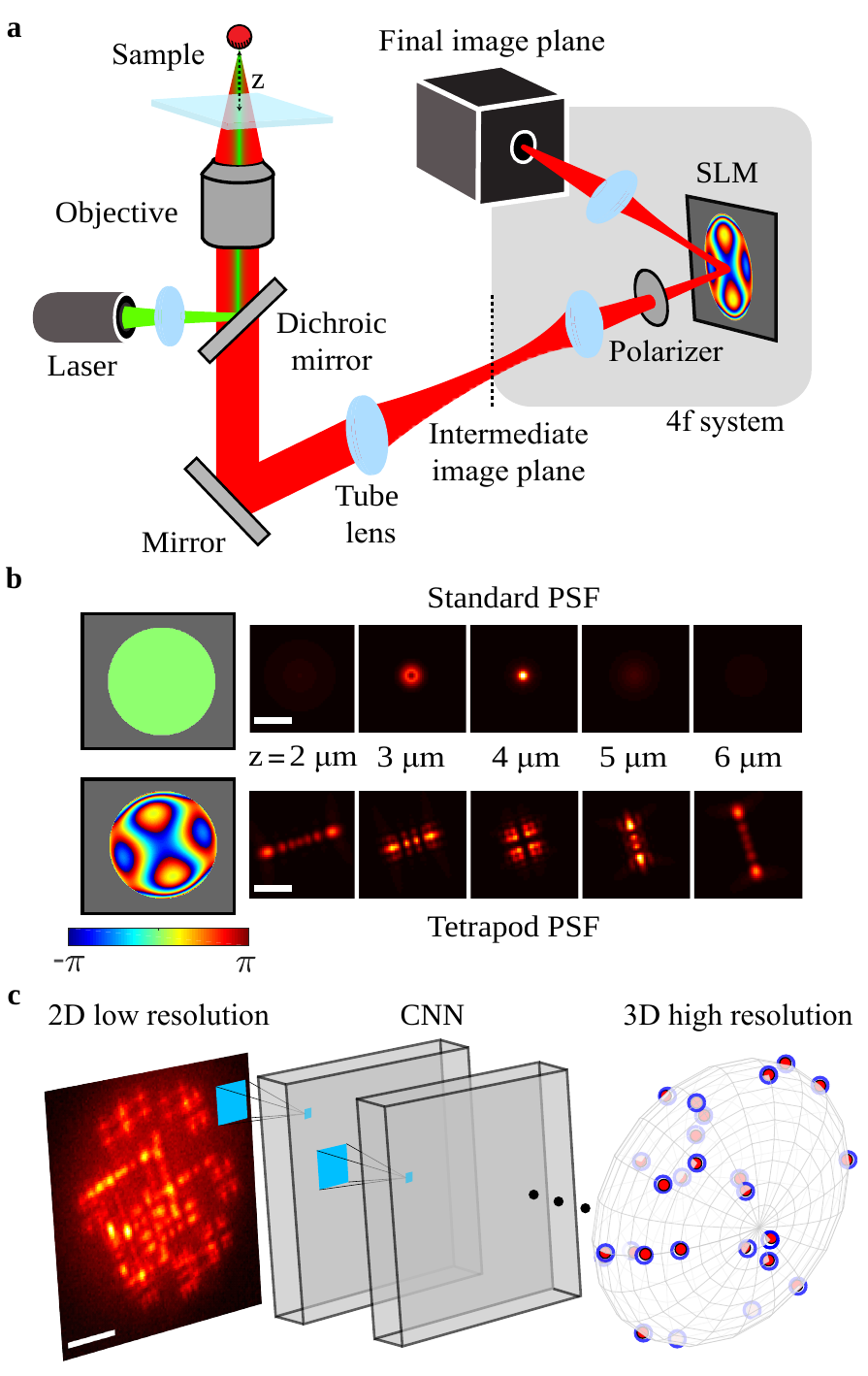}
\caption{\textbf{Optical setup and approach overview}. \textbf{a} The light emitted from a fluorescent microscopic particle is collected by the objective and focused through the tube lens into an image at the intermediate image-plane. This plane is extended using a 4f system with a phase mask placed at the Fourier plane in between the two 4f lenses. \textbf{b} The implemented phase mask  (using either a Spatial Light Modulator (SLM) or fabricated fused-silica) dictates the shape of the PSF as function of the emitter's axial position. \textbf{c} After training, our CNN receives a 2D low resolution image of overlapping PSFs and outputs a 3D high-resolution volume which is translated to a list of 3D localizations. Blue empty spheres denote simulated GT positions along the surface of an ellipsoid. Red spheres denote CNN detections. The Tetrapod PSF is depicted here, however the approach is applicable to any PSF, including those optimized by the net itself (Fig. \ref{fig:psf-learning}). Scale bars are 3 \(\mu\)m.}
\label{fig:concept}
\end{figure}

\begin{figure*}[!ht]
\centering
\includegraphics{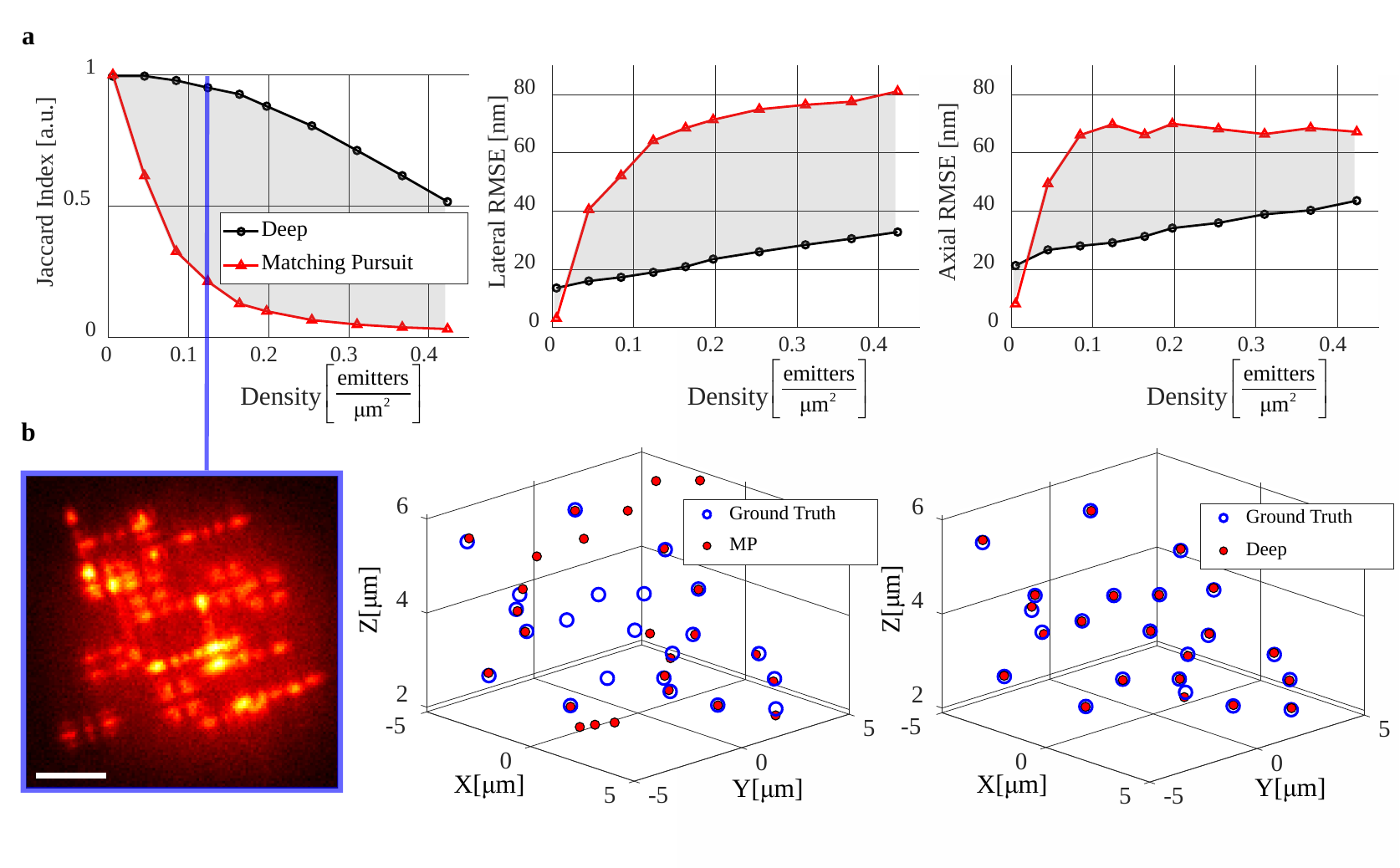}
\caption{\textbf{Comparison to MP}. \textbf{a} The trained CNN is superior to the matching pursuit approach in both detectability (Jaccard index) and in accuracy (Lateral\textbackslash Axial RMSE). Matching of points was computed with a threshold distance of 150 nm using the Hungarian algorithm \citep{kuhn1955hungarian}. \textbf{b} Example of a simulated frame of density 0.124 \(\left[\frac{\text{emitters}}{\mu m^2}\right]\) alongside 3D comparisons of the recovered positions by MP (middle) and by the CNN (right). Scale bar is 2 \(\mu\)m.}
\label{fig:comparison-mp}
\end{figure*}

In a standard microscope, the precise z position of an emitter is difficult to ascertain because the change of the point-spread function (PSF) near the focus is approximately symmetric. Furthermore, outside of this focal range (\(\approx \pm\) 350 nm for a high numerical aperture imaging system), the rapid defocusing of the PSF reduces the signal-to-noise ratio causing the localization precision to quickly degrade. One method to extend the useful z-range and explicitly encode the z position is PSF engineering \citep{pavani2009three, huang2008three,shechtman2014optimal}. Here, an additional optical element, e.g. a phase mask, is placed in the emission path of the microscope, modifying the image formed on the detector \citep{backer2014extending} (Fig. \ref{fig:concept}a); the axial position can then be recovered via image processing using a theoretical or experimentally-calibrated PSF model  \citep{shechtman2014optimal, liu2013three, babcock2017analyzing, li2018real}.

In practically all applications, it is desirable to be able to localize nearby emitters simultaneously. For example, in super-resolution SMLM experiments, the number of emitters localized per frame determines the temporal resolution. In tracking applications, PSF overlap from multiple emitters often precludes localization, potentially biasing results in emitter-dense regions. The problem is that localizing overlapping emitters poses a significant algorithmic challenge even in 2D localization, and much more so in 3D. Specifically, encoding the axial position of an emitter over large axial ranges (>3 \(\mu\)m) requires the use of laterally large PSFs, e.g. the Tetrapod \citep{shechtman2014optimal, shechtman2015precise} (Fig. \ref{fig:concept} b), increasing the possibility of overlap. Consequently, while a variety of methods have been developed to cope with overlapping emitters for the in-focus, standard-PSF \citep{min2014falcon, boyd2017alternating, nehme2018deep}, the performance in high-density 3D localization situations is far from satisfactory \citep{sage2019super}.

Deep learning has proven to be adept at analyzing microscopic microscopy data \citep{rivenson2018phase, nguyen2018deep, weigert2018content, rivenson2019phasestain, liu2019deep, smith2019ultra}, especially for single-molecule localization, handling dense fields of emitters over small axial ranges (<1.5 \(\mu\)m) \citep{nehme2018deep, boyd2018deeploco, ouyang2018deep, diederich2019cellstorm, newby2018convolutional, zelger2018three, liu2018fast,hershko2019multicolor, speiser2019teaching} or sparse emitters spread over larger ranges \citep{zhang2018analyzing}. Moreover, an emerging application is to jointly design the optical system alongside the data processing algorithm, enabling end-to-end optimization of both components \citep{chakrabarti2016learning, horstmeyer2017convolutional, turpin2018light, haim2018depth, he2018learning, hershko2019multicolor, sitzmann2018end, chang2019deep, wuphasecam3d}. Here we present DeepSTORM3D, consisting of two fundamental contributions to high-density 3D localization microscopy over large axial ranges. First, we employ a convolutional neural network (CNN) for analyzing dense fields of overlapping emitters with engineered PSFs, demonstrated with the large-axial-range Tetrapod PSF \citep{shechtman2014optimal, shechtman2015precise}. Second, we design an optimal PSF for 3D localization of dense emitters over a large axial range of 4 \(\mu\)m. By incorporating a physical-simulation layer in the CNN with an adjustable phase modulation, we jointly learn the optimal PSF (encoding) and associated localization algorithm (decoding). This approach is highly flexible and easily adapted for any 3D SMLM dataset parameters, \textit{i.e.} emitter density, SNRs, and z-range. We quantify the performance of the method by simulation, and demonstrate the applicability to 3D biological samples, \textit{i.e.} mitochondria and telomere.

\begin{figure*}[!htb]
\centering
\includegraphics{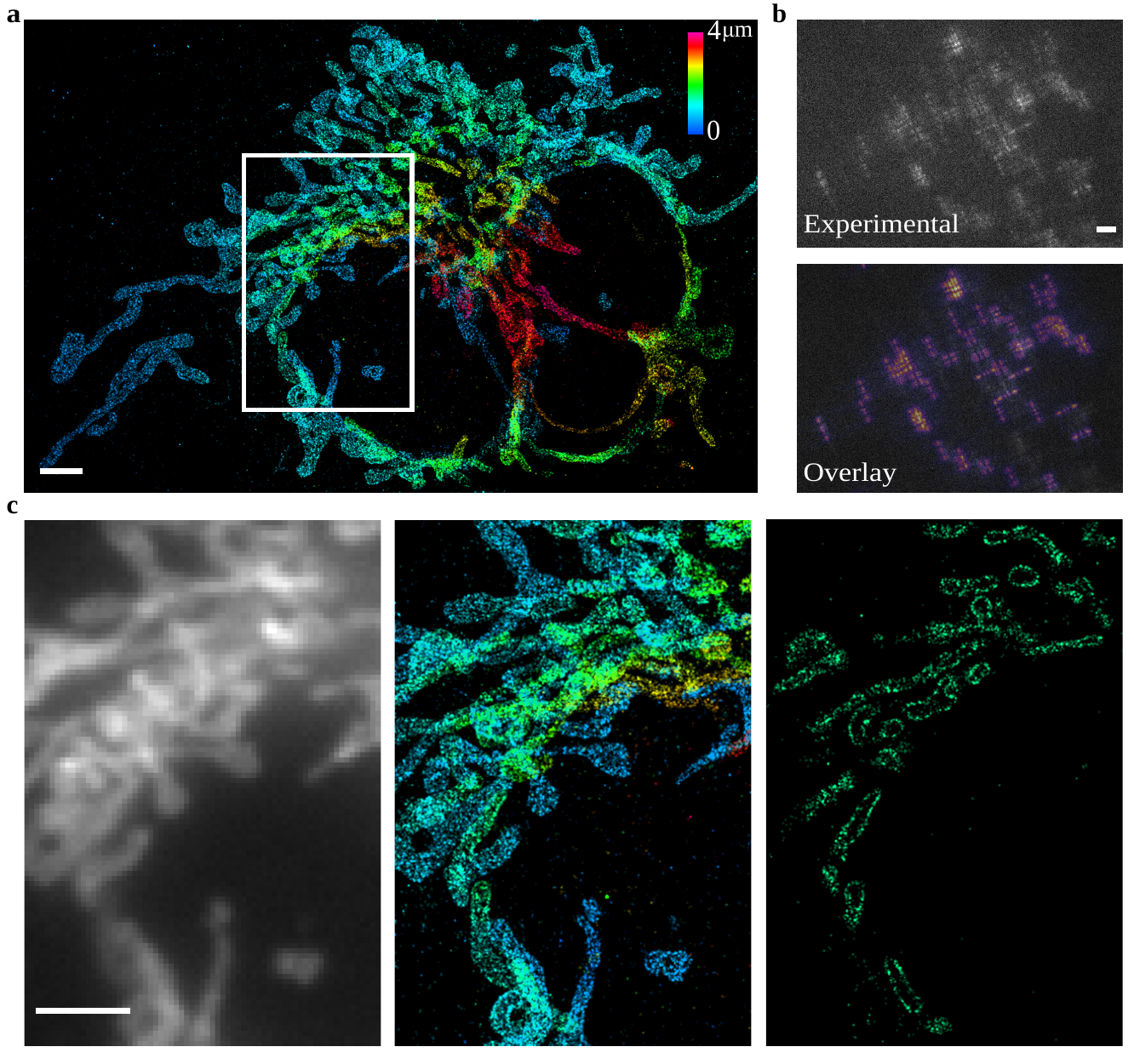}
\caption{\textbf{Super-resolution 3D imaging over a 4 \(\mu\)m z-range}. \textbf{a} Super-resolved image of mitochondria spanning a \(\approx\)4\(\mu\)m z-range rendered as a 2D histogram where z is encoded by color. \textbf{b} Representative experimental frame (top), and rendered frame from the 3D recovered positions by the CNN overlaid on top (bottom). \textbf{c} Diffraction limited (left), super-resolved (middle), and cross-section of the super-resolved image at \(z=1.5 \ \mu\)m (right). Scale bars are 3 \(\mu\)m.}
\label{fig:storm3d}
\end{figure*}

\section{Results}

To solve the high-density localization problem in 3D, we trained a CNN that receives a 2D image of overlapping Tetrapod PSFs spanning an axial range of 4 \(\mu\)m, and outputs a 3D grid with a voxel-size of \(27.5 \times 27.5 \times 33 \ nm^3\) (Fig. \ref{fig:concept}c). For architecture details and learning hyper-parameters see Supplementary Information sections 1.1 and 3. To compile a list of localizations, we apply simple thresholding, and local maximum finding on the output 3D grid (Supplementary Information section 3.4).

\begin{figure*}[!h]
\centering
\includegraphics{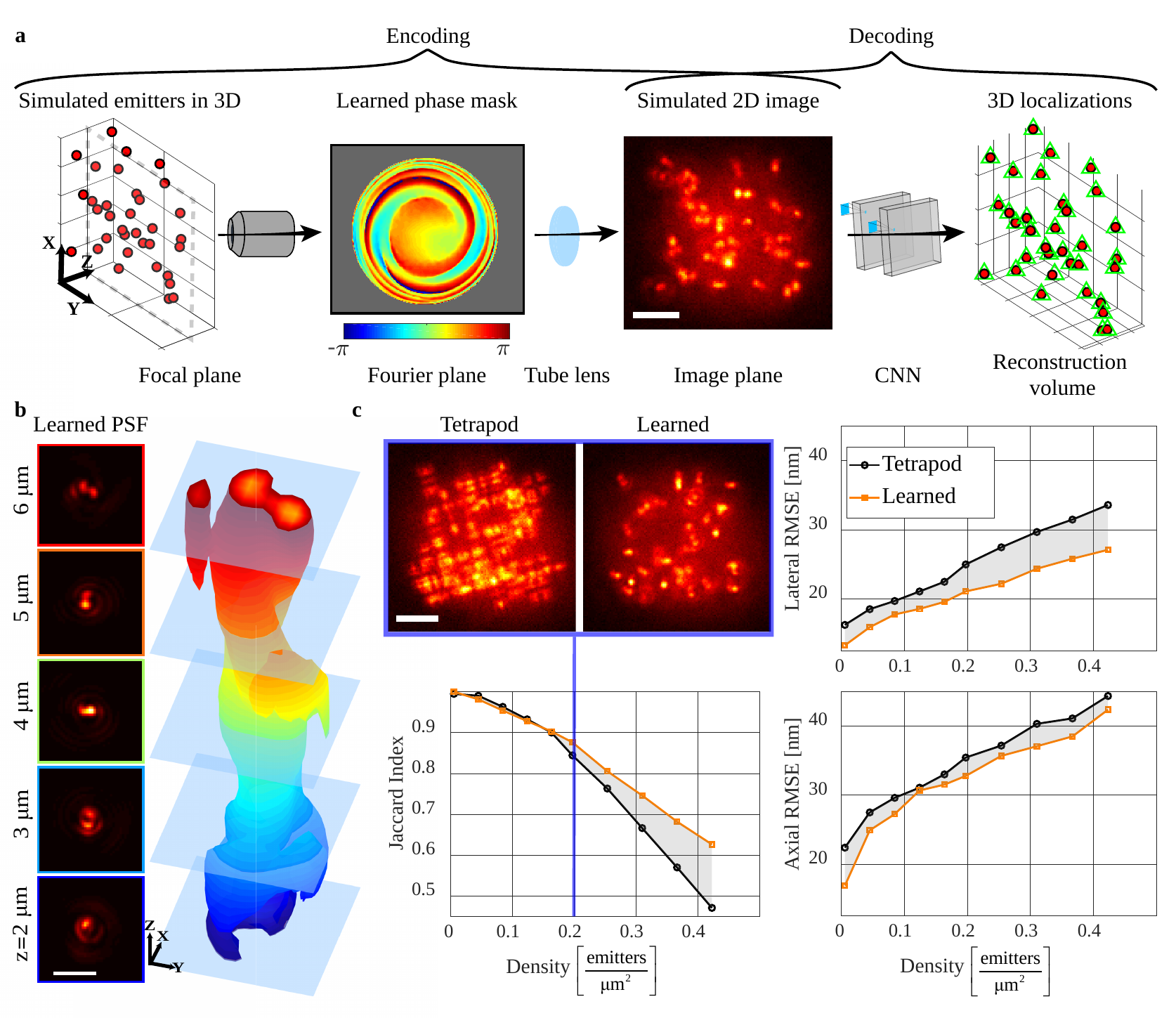}
\caption{\textbf{PSF learning for high density 3D imaging}. \textbf{a} Simulated 3D emitter positions are fed to the image formation model to simulate their low resolution CCD image (Encoding). Next, this image is fed to a CNN that tries to recover the simulated emitter positions (Decoding). The difference between the simulated positions and the positions recovered by the CNN is used to jointly optimize the phase mask at the Fourier plane, and the recovery CNN parameters. \textbf{b} Simulation of the learned PSF as function of the emitter axial position (left). 3D isosurface rendering of the learned PSF (right). \textbf{c} Example frame of density 0.197 \(\left[\frac{\text{emitters}}{\mu m^2}\right]\) (top) with the same simulated emitter positions, using the Tetrapod  (left) and the learned PSF (right). Jaccard index (bottom) and lateral \textbackslash axial RMSE comparison (right) between two CNNs with the same architecture, one trained to recover 3D positions from 2D images of Tetrapod PSF (black), and the second trained to recover 3D positions from 2D images of the learned PSF (orange). Scale bars are 3 \(\mu\)m.}
\label{fig:psf-learning}
\end{figure*}

\begin{figure*}[!ht]
\centering
\includegraphics{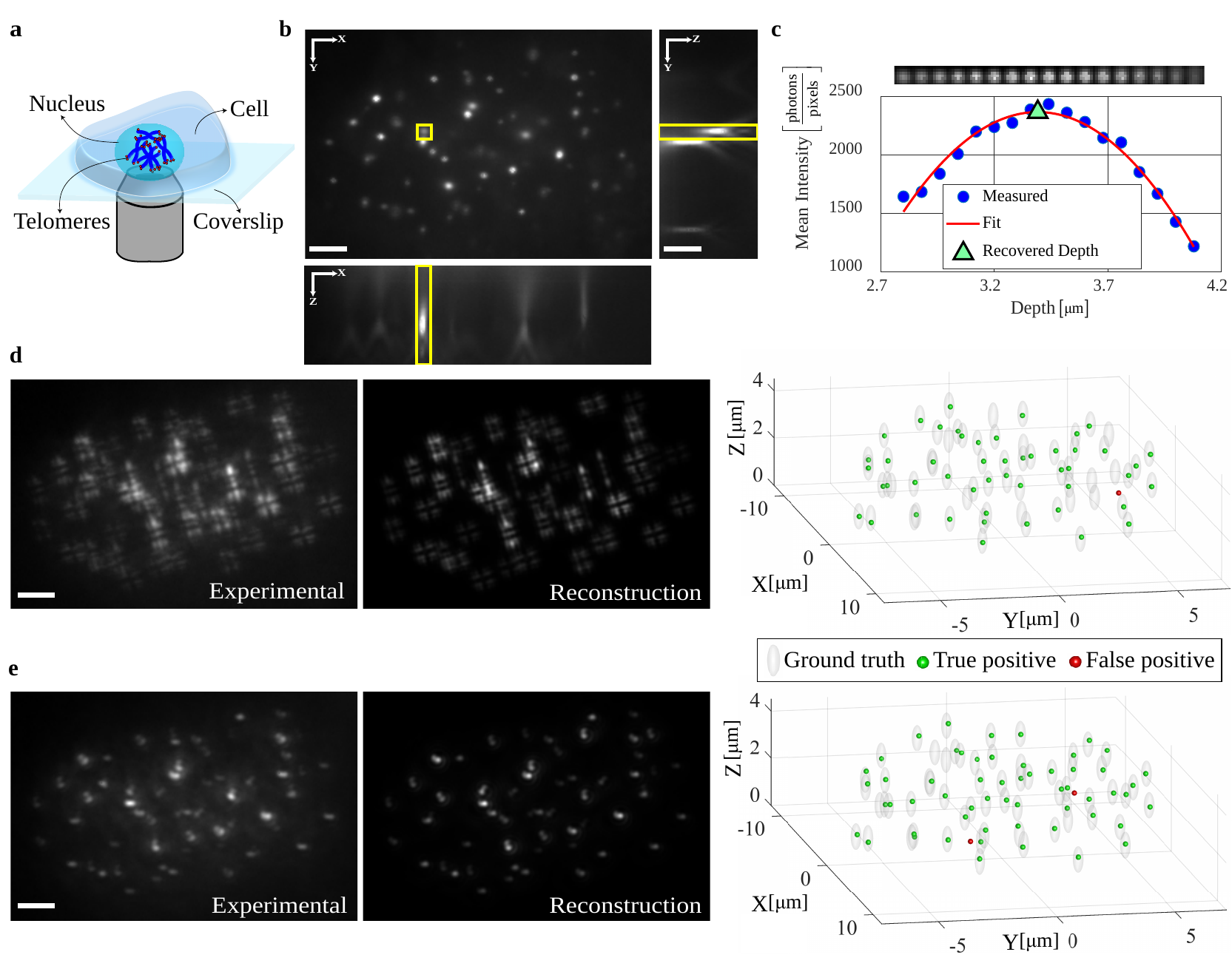}
\caption{\textbf{Three dimensional imaging of telomeres in a single-snapshot}. \textbf{a} Schematic of imaging fixed U2OS cells with fluorescent labeled telomeres inside their nucleus. \textbf{b} Focus slice with the standard PSF inside a U2OS cell nucleus, obtained via a z-scan. The yellow rectangles mark the same emitter in all three orthogonal planes. \textbf{c} Example fit of the mean intensity in sequential axial slices used to estimate the approximate emitter axial position. \textbf{d} Experimental snapshot with the Tetrapod PSF (left), rendered image from the 3D recovered positions by the Tetrapod CNN (middle), and a 3D comparison of the recovered positions and the approximate experimental ground truth (right). \textbf{e} Experimental snapshot with the learned PSF (left), rendered image from the 3D recovered positions by the learned PSF CNN (middle), and a 3D comparison of the recovered positions and the approximate experimental ground truth (right). Scale bars are 3 \(\mu\)m.}
\label{fig:exp-demo}
\end{figure*}

We compare our method to a fit-and-subtract based Matching Pursuit (MP) approach \citep{shechtman2016multicolour} (see Supplementary Information section 4) as we are unaware of any other methods capable of localizing overlapping Tetrapod PSFs. To quantitatively compare our method with MP solely in terms of density, we simulated emitters with high signal-to-noise ratio (30K signal photons, 150 background photons per pixel) at 10 different densities ranging from 1 to 75 emitters per 13 \(\times\) 13   \(\mu m^2\) field-of-view. The results are shown in Fig. \ref{fig:comparison-mp}. As evident in both the Jaccard index (defined as \(\frac{TP}{TP + FP + FN}\), where TP, FP, FN are true positives, false positives, and false negatives \citep{sage2019super}) and the lateral/axial RMSE (Fig. \ref{fig:comparison-mp}a) the CNN achieves remarkable performance in localizing high-density Tetrapods. In the single-emitter (very low density) case, where the performance of the CNN is bounded by the discretization on the 3D grid, the RMSE of the MP localization is lower (better). This is because for a single-emitter, MP is equivalent to a continuous Maximum Likelihood Estimator (MLE) (Supplementary Information section 4), which is asymptotically optimal \citep{bickel2015mathematical}, whereas the CNN's precision is bounded by pixilation of the grid (\textit{i.e.} half voxel of 13.75 nm in xy and 16.5 nm in z). However, quickly beyond the single-emitter case, the CNN drastically outperforms MP. A similar result was obtained when compared to a leading single-emitter fitting method \citep{li2018real} applicable also for the multiple emitter case \citep{sage2019super} (see Supplementary Information section 6).

Next, we validated our method for super-resolution imaging of fluorescently labeled mitochondria in COS7 cells (Fig. \ref{fig:storm3d}). We acquired 20K diffraction limited frames of a \(50\times30 \ \mu m^2\) FOV and localized them using the CNN in about \(\approx\)10 hours, resulting in \(\approx\)360K localizations. The Tetrapod PSF was implemented using a fabricated fused-silica phase-mask (see Supplementary Information section 7.1). The estimated resolution was \(\approx\)40 nm in xy, and \(\approx\) 50 nm in z (see Supplementary Information section 7.2). To visually evaluate localization performance in a single frame (Fig. \ref{fig:storm3d}b top), we regenerated the corresponding 2D low-resolution image, and overlayed the recovered image with a uniform photons scaling on top of the experimental frame \ref{fig:storm3d}b bottom). As seen in the overlay image, the emitter PSFs (3D positions) are faithfully recovered by the CNN. Moreover, emitters with extremely low number of signal photons were ignored.

The Tetrapod is a special PSF that has been optimized for the single emitter case by Fisher Information maximization \citep{shechtman2014optimal,shechtman2015precise}. However, when considering the multiple-emitter case, an intriguing question arises: What is the optimal PSF for high density 3D localization over a large axial range? To answer this question we need to rethink the design metric; extending the Fisher Information criterion \citep{shechtman2014optimal} to account for emitter density is not-trivial, and while it is intuitive that a smaller-footprint PSF would be preferable for dense emitters, it is not clear how to mathematically balance this demand with the requirement for high localization precision per emitter.

Our PSF-design logic is based on the following: since we have already established that a CNN yields superior reconstruction for high-density 3D localization, \textit{we are interested in a PSF (\textit{encoder}) that would be optimally localized by a CNN (\textit{decoder})}. Therefore, in contrast to a sequential paradigm where the PSF and the localization algorithm are optimized separately, we adopt a co-design approach (Fig. \ref{fig:psf-learning}a). To jointly optimize the PSF and the localization CNN, we introduce a differentiable physical simulation layer, which is parametrized by a phase mask that dictates the microscope's PSF. This layer encodes 3D point sources to their respective low-resolution 2D image (see Supplementary Information section 2). This image is then fed to the localization CNN which decodes it and recovers the underlying 3D source positions. During training, the net is presented with simulated point sources at random locations and, using the difference between the CNN recovery and the simulated 3D positions, we optimize both the phase mask and the localization CNN parameters in an end-to-end fashion. The learned PSF (Fig. \ref{fig:psf-learning}b) has a small lateral footprint, which is critical for minimizing overlap at high densities. Moreover, the learned phase mask twists in a spiral trajectory causing the PSF to rapidly rotate throughout the axial range, a trait that was previously shown to be valuable for encoding depth \citep{pavani2009three}.

To quantify the improvement introduced by our new PSF, we first compare it to the Tetrapod PSF in simulations. Specifically, we train a similar reconstruction net for both the Tetrapod and the learned PSF using a matching training set composed of simulated continuous 3D positions along with their corresponding 2D low-resolution images. The learned PSF performs similar to the Tetrapod PSF for low emitter densities (Fig. \ref{fig:psf-learning}c). However, as the density goes up the learned PSF outperforms the Tetrapod PSF in both localization precision and in emitter detectability (Jaccard index) (Fig. \ref{fig:psf-learning}c). This result is not surprising, as the learned PSF has a smaller spatial foorprint, and hence it is less likely to overlap than the Tetrapod (Fig. \ref{fig:psf-learning}c).

Next, we demonstrate the superiority of the new PSF experimentally by imaging fluorescently labeled telomeres (TRF1-DsRed) in Fixed U2OS cells. The cell contains tens of telomeres squeezed in the volume of a nucleus with \(\approx 20 \ \mu\)m diameter (Fig. \ref{fig:exp-demo}a, b). From a single snapshot focused inside the nucleus, the CNN outputs a list of 3D positions of telomeres spanning an axial range of \(\approx 3 \ \mu\)m. Using the Tetrapod PSF snapshot, the Tetrapod-trained CNN was able to recover 49 out of 62 telomeres with a single false positive, yielding a Jaccard index of 0.77 (Fig. \ref{fig:exp-demo}d). In comparison, using the learned PSF snapshot, the corresponding CNN was able to recover 57 out of the 62 telomeres with only 2 false positives, yielding a Jaccard index of 0.89 (Fig. \ref{fig:exp-demo}e). The recovered positions were compared to approximated ground-truth 3D positions (Fig. \ref{fig:exp-demo}c), obtained by axial scanning and 3D fitting (see Supplementary Information section 9).

To qualitatively compare the recovered list of localizations to the acquired snapshot, we fed this list to the physical simulation layer and generated the matching 2D low-resolution image (Fig. \ref{fig:exp-demo}d,e). As verified by the regenerated images, the 3D positions of the telomeres are faithfully recovered by the CNNs. Moreover, the misses in both snapshots were either due to local aberrations and/or an extremely low number of signal photons (see Supplementary Information section 10 for more experimental results).

\section{Discussion}

In this work we demonstrated 3D localization of dense emitters over a large axial range both numerically and experimentally. The described network architecture exhibits excellent flexibility in dealing with various experimental challenges, e.g. low signal-to-noise ratios and optical aberrations.  This versatility is facilitated in three ways: (1) the net was trained solely on simulated data, thus producing sufficiently large datasets for optimization; (2) the phase mask which governs the PSF was optimized with respect to the implementation in the imaging system, \textit{i.e.} the pixels of the spatial light modular, rather than over a smaller subspace, e.g. Zernike polynomials \citep{shechtman2014optimal}; (3) the CNN localization algorithm was designed in coordination with the development of the PSF, thus the system was optimized for the desired output \citep{hershko2019multicolor} rather than a proxy.

Attaining a sufficiently large training dataset has thus far been a major limitation for most applications of CNNs. With this limitation in mind, the application of CNNs to single-molecule localization would seemingly be an ideal one, since each emitter's behavior should be approximately the same. This uniformity is broken, however, by spatially-varying background, sample density, and variable emitter size in biological samples (Supplementary Information 3.1), all of which diversify datasets and necessitate relevant training data. By implementing an accurate simulator, we have shown that it is possible to build a robust network entirely in silico, generating arbitrarily large, realistic datasets with a known ground truth to optimize the nets. This aligns with our previous work in 2D SMLM \citep{nehme2018deep}).

For super-resolution reconstructions using the Tetrapod PSF, the simulator was particularly important due to the highly variable SNR of emitters in the sample. Here, our net was able to selectively localize the emitters even in very dense regions by focusing on those with a high SNR (Fig. \ref{fig:storm3d}). To optimize a PSF while simultaneously training the net, the simulator was also essential, as it would be prohibitively time consuming to experimentally vary the PSF, while recording and analyzing images to train the net.

An intriguing aspect of our optimization approach is that the optimized PSF is found by continuously varying the pixels of an initialized mask while evaluating the output of the localization net, thus the final result represents a local minimum (Fig. \ref{fig:psf-learning}). By changing the initialization conditions, we have recognized several patterns that indicate how the optimal PSF varies with the experimental conditions: namely, density, axial range, and SNR (see Supplementary Information section 1.2). Some of the recurrent features are intuitive: for example, in dense fields of emitters with limited SNR, the optimized PSFs have a small footprint over the designed axial range, enabling high density and compacting precious signal photons into as few pixels as possible. What distinguishes the net PSFs over predetermined designs is the utilization of multiple types of depth encoding; namely, simultaneously employing astigmatism, rotation, and side lobe movement (Fig. \ref{fig:psf-learning}), all of which have been conceived of and implemented previously, but never simultaneously!

This work, therefore, triggers many possible questions and research directions regarding its capabilities and limitations. For example, how globally-optimal is the resulting PSF? Similarly, how sensitive is the resulting PSF and its performance to different loss functions, CNN architectures, initializations (e.g. with an existing phase mask), and the sampled training set of locations? Currently, it is unclear how each of these components affects the learning process although we began to partially answer them in simulations (see Supplementary Information section 1.2). Finally, the co-design approach employed here paves the way to a wide variety of interesting applications in microscopy where imaging systems have traditionally been designed separately from the processing algorithm.

\section*{Funding Information}

Google; H2020 European Research Council Horizon 2020 (802567); Israel Science Foundation (ISF) (852/17); Ollendorff Foundation; Technion-Israel Institute of Technology (Career Advancement Chairship); Zuckerman Foundation.

\section*{Acknowledgements}
The authors wish to thank the Garini lab for the U2OS cells. We also thank Jonas Ries for his help with the application of SMAP-2018 to Tetrapod PSFs. We gratefully acknowledge the support of NVIDIA Corporation with the donation of the Titan V GPU used for this research. We thank the staff of the Micro-Nano-Fabrication \(\&\) Printing Unit (MNF-PU) at the Technion for their assist with the phase mask fabrication. Finally, we thank Google for the research cloud units provided to accelerate this research.

\section*{Author contributions}
EN, DF, TM and YS conceived the approach. EN performed the simulations and analyzed the data with contributions from all authors. EN, RG, BF, LEW, and OA took the data. RO fabricated the physical phase mask. EN, DF, LEW, TM and YS wrote the paper with contributions from all authors.

\section*{Competing Interests}
The authors declare no competing interests.

\section*{Data availability}
Data will be made available upon reasonable request.

\section*{Code availability}
Code will be made publicly available.

\bibliographystyle{ieeetr}
\bibliography{main_new}

\section*{Methods}

\subsection*{Sample preparation}

COS7 cells were grown for 24 hr on cleaned \(22 \times 22\) mm, 170 \(\mu\)m thick coverslips in 6-well plate in Dulbecco's Modified Eagle Medium (DMEM) With 1g/l D-Glucose (Low Glucose), supplemented with Fetal bovine serum, Penicillin-Streptomycin and glutamine, at \(37^o C\), and 5\(\% CO_2\). The cells were fixed with 4\(\%\) paraformaldehyde and 0.2\(\%\) glutaraldehyde in PBS, pH 6.2, for 45 min, washed and incubated in 0.3M glycine/PBS solution for 10 minutes. The coverslips were transferred into a clean 6-well plate and incubated in a blocking solution for 2 hr (10\(\%\) goat serum, 3\(\%\) BSA, 2.2\(\%\) glycine, and 0.1\(\%\) Triton-X in PBS, filtered with 0.45 \(\mu\)m PVDF filter unit, Millex). The cells were then immunostained overnight with anti TOMM20-AF647 (Abcam , ab209606) 1:230 diluted in the blocking buffer, and washed X5 with PBS. Cover glasses (\(22\times 22\) mm, 170 \(\mu\)m thick) were cleaned in an ultrasonic bath with 5\(\%\) Decon90 at \(60^o\)C for 30 min, then washed with water, incubated in ethanol absolute for 30 min and sterilized with 70\(\%\) filtered ethanol for 30 min.

U2OS cells were grown on cleaned 0.18 \(\mu\)m coverslips in 12-well plate in Dulbecco's Modified Eagle Medium (DMEM) With 1g/l D-Glucose (Low Glucose), supplemented with Fetal bovine serum, Penicillin-Streptomycin and glutamine, at \(37^o C\), and 5\(\% CO_2\). The day after cells were transfected with dsRed-TRF1 plasmid \citep{bronshtein2015loss} using Lipofectamine 3000 reagent. 24 hr after transfection cells were fixed with 4\(\%\) paraformaldehyde for 20 minutes, washed 3 times with PBS and attached to a slide together with mounting medium.

\subsection*{Optical setup}

Imaging experiments were performed on the experimental system shown schematically in Fig. \ref{fig:concept}b. The 4f optical processing system was built alongside the side-port of a Nikon Eclipse Ti inverted fluorescence microscope, with a \(\times\)100/1.45 numerical aperture (NA) oil-immersion objective lens (Plan Apo \(\times\)100/1.45 NA, Nikon).

\subsection*{STORM imaging}

For super-resolution imaging, a PDMS chamber was attached to a glass coverslip containing fixed COS7 cells. Blinking buffer (100 mM \(\beta\)-mercaptoethylamine hydrochloride, 20\(\%\) sodium lactate, and 3\(\%\) OxyFluor (Sigma, SAE0059), modified from \citep{nahidiazar2016optimizing}, was then added and a glass coverslip was placed on top to prevent evaporation.
Low-intensity illumination for recording diffraction-limited images was applied using a Topica laser (640 nm), on the Nikon TI imaging setup described previously, and recorded with an EMCCD (iXon, Andor) in a standard imaging setup. For super-resolution blinking using the Tetrapod PSF, high-intensity (1W at the back of the objective lens) 640 nm light was applied using a 638 nm 2000 mW red dot laser module, whose beam shape was cleaned using a 25 \(\mu\)m pinhole (Thorlabs) in coordination with low-intensity (<5mW) 405 nm light. Emission light was filtered through a 500 nm Long pass dichroic and a 650 nm long pass (Chroma), projected through a 4f system containing the dielectric Tetrapod phase mask (see Supplementary Information section 7.1), and imaged on a Prime95b Photometrics camera. 

\subsection*{Super-resolution image rendering}

Prior to rendering the super-resolved image (Fig. \ref{fig:storm3d}b), we first corrected for sample drift using the ThunderSTORM \citep{ovesny2014thunderstorm} ImageJ \citep{schindelin2012fiji} plugin. Afterwards, we rendered the 3D localizations as a 2D average shifted histogram, with color encoding the z position.

\subsection*{Telomere imaging}

For telomere imaging, the 4f system consisted of two f=15 cm lenses (Thorlabs), a linear polarizer (Thorlabs) to filter out the light that is polarized in the unmodulated direction of the SLM, a \(1920\times1080\) pixel SLM (PLUTO-VIS, Holoeye) and a mirror for beam-steering. A sCMOS camera (Prime95B, Photometrics) was used to record the data. The sample was illuminated with 561 nm fiber-coupled laser-light source (iChrome MLE, Toptica). The excitation light was reflected up through the microscope objective by a multibandpass dichroic filter (TRF89902-EM - ET - 405/488/561/647nm Laser Quad Band Set, Chroma). Emission light was filtered by the same dichroic and also filtered by another 617 nm band pass filter (FF02-617/73 Semrock).

\subsection*{CNN Architecture}

In a nutshell, our localization CNN architecture is composed of 3 main modules. First, a multi-scale context aggregation module process the input 2D low-resolution image and extracts features with a growing receptive field using dilated convolutions \citep{yu2015multi}. Second, an up-sampling module increase the lateral resolution of the predicted volume by a factor of x4. Finally, the last module refines the depth and lateral position of the emitters and outputs the predicted vacancy-grid. For more details regarding the architecture see Supplementary Information section 1.

\end{document}